\documentclass[12pt,cite,epsf,epsfig]{article}
\usepackage{epsfig}

\begin{document}

\author{S. Dev\thanks{dev5703@yahoo.com}, Shivani Gupta\thanks{shiroberts\_1980@yahoo.co.in} and Radha Raman Gautam\thanks{gautamrrg@gmail.com}}

\title{Zero Textures of the Neutrino Mass Matrix from Cyclic Family Symmetry}
\date{\textit{Department of Physics, Himachal Pradesh University, Shimla 171005, India.}\\
\smallskip}

\maketitle
\begin{abstract}
We present the symmetry realization of the phenomenologically viable Frampton-Glashow-Marfatia (FGM) two zero texture neutrino mass matrices in the flavor basis within the framework of the type (I+II) seesaw mechanism natural to SO(10) grand unification. A small Abelian cyclic symmetry group $Z_3$ is used to realize these textures except for class C for which the symmetry is enlarged to $Z_4$. The scalar sector is restricted to the Standard Model (SM) Higgs doublet to suppress the flavor changing neutral currents. Other scalar fields used for symmetry realization are at the most two scalar triplets and, in some cases, a complex scalar singlet. Symmetry realization of one zero textures has, also, been presented.
\end{abstract}

\section{Introduction}
Understanding the pattern of neutrino masses and mixings constitutes a major challenge for elementary particle physics. This pattern seems to be entirely different from the observed hierarchical pattern of quark masses and mixings. In the absence of flavor symmetries, fermion masses and mixings are, in general, undetermined in gauge theories. All information about fermion masses and mixings is encoded in the relevant fermion mass matrices which are important tools for the investigation of the underlying symmetries and the resulting dynamics. The important first step in this direction is the reconstruction of the neutrino mass matrix in the flavor basis. However, this reconstruction results in a large variety of possible structures of the neutrino mass matrix depending strongly on the neutrino mass scale, mass hierarchy and the $CP$ violating phases. In fact, no presently conceivable set of feasible experiments can determine the neutrino mass matrix completely. Therefore, in the absence of sufficient data on neutrino masses and mixings, all possible mass matrix structures need to be carefully scrutinized to find viable structures compatible with the presently available data. Several interesting proposals have been made in the literature to restrict the form of the neutrino mass matrix and, thus, reduce the number of free parameters. These include the presence of zero textures \cite{1,2,3}, vanishing minors \cite{4} and hybrid textures \cite{5} to name just a few. Zero textures in the neutrino mass matrix, in particular, have been extensively studied in the literature. The main reason for such interest in zero textures is their implications for the possible existence of family symmetries which require certain entries of the mass matrix to vanish. It has been noted earlier \cite{6} that the zeros of the Dirac neutrino mass matrix $(M_D)$ and the heavy right-handed Majorana neutrino mass matrix $(M_R)$ may propagate as zeros in the effective low energy neutrino mass matrix $(M_\nu)$ through the seesaw mechanism \cite{7}. Zero textures in the neutrino mass matrix, in general, may be obtained by imposing certain Abelian family symmetries \cite{8} at the expense of an extended scalar sector. Abelian family symmetries have been investigated systematically in \cite{9} for extremal mixings. General guidelines for the symmetry realization of zero textures in both the quark and the lepton mass matrices have been propounded in \cite{10} which outline general methods for enforcing zero entries in the fermion mass matrices by imposing Abelian family symmetries. However, specific guidelines for a simple and minimal realization of each texture are still lacking.\\ Two zero texture Ans\"{a}tz is especially important since it can successfully describe both the quark and lepton sectors including the charged lepton and the neutrino masses. The two zero textures are, also, compatible with specific GUT models \cite{11}. Furthermore, these mass matrices can accommodate the present values of sin$2\beta$ \cite{12}. In view of the phenomenological success of the two zero texture Ans\"{a}tz, it would be interesting to examine the symmetry realization of two zero texture neutrino mass matrices in the larger context of SO(10) GUT which includes both the type I and type II contributions in the seesaw mechanism. Out of the seven phenomenologically viable classes of two zero texture neutrino mass matrices (FGM), two classes have been obtained in \cite{13} from $A_4$ or its $Z_3$ subgroup in the context of type (I+II) seesaw \cite{15}. Symmetry realization of all the phenomenologically viable classes of two zero texture neutrino mass matrices was presented in \cite{14} in the context of type II seesaw mechanism. In the present work, we present the symmetry realization of all the phenomenologically viable classes of two zero texture neutrino mass matrices in the context of type (I+II) seesaw \cite{15} without assuming the dominance of either type of contributions. We keep just the Standard Model (SM) scalar doublet which transforms trivially under the new family symmetry ($Z_3/Z_4$) thus, in effect, suppressing the undesired flavor changing neutral currents. An additional advantage of a single scalar doublet is the stability of zero textures in the neutrino mass matrix under renormalization group evolution. The neutrino mass matrices at any two scales $\mu_1$ and $\mu_2$ are related by \cite{14, 16} $M_\nu(\mu_1)=IM_{\nu}(\mu_2)I$, where $I$ is diagonal and positive leading to zeros in $M_{\nu}$ at any other renormalizable scale. These zero textures are enforced by extending the SM with at the most two scalar $SU(2)$ triplets. However, in some cases we are forced to introduce a complex scalar singlet which transforms nontrivially under the family symmetry. The effective neutrino mass matrix contains both type (I+II) seesaw contributions which is natural in SO(10) GUTs as
\begin{equation}
M_{\nu}=M_{\nu}^{I} +M_{\nu}^{II}=M_L-M_DM_R^{-1}M_D^T
\end{equation}
where $M_L$ is the left-handed Majorana mass matrix. Symmetry realization of one zero texture neutrino mass matrices has, also, been discussed briefly.
\section{Symmetry Realization of Two Zero Textures}
The phenomenologically allowed two zero texture neutrino mass matrices  are given in Table 1. 
\begin{table}
\begin{center}
\begin{tabular}{|c|c|c|c|}
\hline $A_1$ & $A_2$ &$B_1$ & $B_2$ \\ 
 \hline $\left(
\begin{array}{ccc}
0 & 0 & X \\ 0 &X & X \\ X& X & X
\end{array}
\right)$ &$\left(
\begin{array}{ccc}
0 & X & 0 \\ X &X & X \\ 0& X & X
\end{array}
\right)$&$\left(
\begin{array}{ccc}
X & X & 0 \\ X &0 & X \\ 0& X & X
\end{array}
\right)$ & $\left(
\begin{array}{ccc}
X & 0 & X \\ 0 &X & X \\ X& X & 0
\end{array}
\right)$  \\ 
\hline $B_3$ & $B_4$ &$C$&-\\ 
\hline $\left(
\begin{array}{ccc}
X & 0 & X \\ 0 &0 & X \\X& X & X
\end{array}
\right)$ & $\left(
\begin{array}{ccc}
X & X & 0 \\ X &X & X \\ 0& X & 0
\end{array}
\right)$&$\left(
\begin{array}{ccc}
X & X & X \\ X &0 & X \\X& X & 0
\end{array}
\right)$&-   \\ \hline
\end{tabular}
\caption{Viable two zero texture neutrino mass matrices. X denote the non zero elements.}
\end{center}
\end{table}
The cyclic symmetry $Z_3$ can be used to realize class $A_1$ when the leptonic fields transform as
\begin{eqnarray}
 D_{L_e}\rightarrow  \omega^2 D_{L_e},& e_{R}\rightarrow \omega^2 e_{R},&  \nu_{R_1}\rightarrow \omega \nu_{R_1},  \nonumber \\ D_{L_\mu}\rightarrow  D_{L_\mu},& \mu_{R}\rightarrow \mu_{R}, & \nu_{R_2}\rightarrow  \nu_{R_2},  \\ D_{L_\tau}\rightarrow \omega D_{L_\tau},& \tau_{R}\rightarrow  \omega \tau _{R}, & \nu_{R_3}\rightarrow \nu_{R_3}, \nonumber
\end{eqnarray}
where $\omega$ = exp($2i \pi/3$) is the generator of $Z_3$.
These assigned transformations generate the diagonal charged lepton mass matrix $M_l$. The bilinears $\overline{D}_{L_j}\nu_{R_k}$ and $\nu_{R_j}\nu_{R_k}$ relevant for $M_D$ and $M_R$ transform as
\begin{center}
$\overline{D}_{L_j}\nu_{R_k} \sim \left(
\begin{array}{ccc}
\omega^2& \omega &\omega \\
\omega &1 &1 \\
1& \omega^2& \omega^2
\end{array}
\right)$,
$\nu_{R_j}\nu_{R_k} \sim \left(
\begin{array}{ccc}
\omega^2& \omega &\omega \\
\omega & 1 &1 \\
\omega& 1& 1
\end{array}
\right)$.
\end{center}
Since, the SM Higgs doublet transforms trivially under $Z_3$, the form of $M_D$ becomes
\begin{center}
$M_D= \left(
\begin{array}{ccc}
0& 0 &0 \\
0 &X &X \\
X& 0& 0
\end{array}
\right)$.
\end{center}
Assuming a complex scalar singlet $\chi$ that transforms under $Z_3$ as $\chi \rightarrow \omega^2 \chi$, the form of $M_R$ becomes
\begin{center}
$M_R= \left(
\begin{array}{ccc}
0& X &X \\
X &X &X \\
X& X& X
\end{array}
\right)$.
\end{center}
After the type I seesaw we get
\begin{eqnarray} 
M_{\nu}^I= \left(
\begin{array}{ccc}
0& 0 &0 \\
0 &X &X \\
0& X& X
\end{array}
\right).
\end{eqnarray}
The bilinear $D_{L_j}^T C^{-1}D_{L_k}$ relevant for $M_L$ transforms under this symmetry as
\begin{center}
$D_{L_j}^T C^{-1}D_{L_k} \sim \left(
\begin{array}{ccc}
\omega& \omega^2 &1 \\
\omega^2 &1 &\omega \\
1& \omega& \omega^2
\end{array}
\right)$.
\end{center}
We introduce a scalar SU(2) triplet $\Delta$ written in $2\times2$ matrix notation as 
\begin{eqnarray}
\Delta= \left(
\begin{array}{cc}
H^+& \sqrt{2}H^{++}  \\
\sqrt{2}H^{0} &-H^+
\end{array}
\right)
\end{eqnarray}
which remains invariant under $Z_3$. The vacuum expectation value (VEV) of Higgs triplet is
\begin{eqnarray}
\langle\Delta\rangle_0= \left(
\begin{array}{cc}
0& 0 \\
v_t &0
\end{array}
\right)
\end{eqnarray}
where $\langle H^0 \rangle_0=\frac{v_t}{\sqrt{2}}$. This induced VEV in the scalar potential is suppressed by the high mass of the scalar triplet \cite{14, 17}. 
Thus, the type II seesaw contribution is given by
\begin{eqnarray}
M_{\nu}^{II}= \left(
\begin{array}{ccc}
0& 0 &X \\
0 &X &0 \\
X& 0& 0
\end{array}
\right).
\end{eqnarray}
The effective neutrino mass matrix $M_{\nu}$ after type (I+II) seesaw mechanism becomes
\begin{eqnarray}
M_{\nu}= \left(
\begin{array}{ccc}
0& 0 &X \\
0 &X &X \\
X& X& X
\end{array}
\right)
\end{eqnarray}
which is the class $A_1$ of FGM \cite{1} two zero textures.\\
For class $A_2$, the leptonic fields under $Z_3$ transform as
\begin{eqnarray}
 D_{L_e}\rightarrow  \omega^2 D_{L_e},& e_{R}\rightarrow \omega^2 e_{R},&  \nu_{R_1}\rightarrow \omega \nu_{R_1},  \nonumber \\ D_{L_\mu}\rightarrow  \omega D_{L_\mu},& \mu_{R}\rightarrow \omega \mu_{R}, & \nu_{R_2}\rightarrow  \nu_{R_2},  \\ D_{L_\tau}\rightarrow  D_{L_\tau},& \tau_{R}\rightarrow \tau _{R}, & \nu_{R_3}\rightarrow \nu_{R_3}. \nonumber
\end{eqnarray}
In this case, the complex scalar singlet $\chi$ is assumed to transform as $\chi \rightarrow \omega^2 \chi$ under $Z_3$ resulting in the same structure of the right-handed Majorana mass matrix as of $A_1$ and we obtain the same contribution from type I seesaw as in Eqn. (3). Assuming the scalar triplet to remain invariant under $Z_3$, the above given transformations of the left-handed fields yield the type II seesaw contribution given by
\begin{eqnarray} 
M_{\nu}^{II}= \left(
\begin{array}{ccc}
0& X &0 \\
X &0 &0 \\
0&0& X
\end{array}
\right).
\end{eqnarray}
Thus, type (I+II) seesaw results in class $A_2$ of FGM two zero textures. This Fritzsch-type neutrino mass matrix with a nearly diagonal charged lepton mass matrix has been obtained recently in \cite{18} using $S_3 \times Z_5 \times Z_2$ symmetry and an extended Higgs sector.\\
The requisite transformations of the leptonic fields for class $B_3$ are given by
\begin{eqnarray}
 D_{L_e}\rightarrow D_{L_e},& e_{R}\rightarrow e_{R},&  \nu_{R_1}\rightarrow \omega \nu_{R_1},  \nonumber \\ D_{L_\mu}\rightarrow \omega^2 D_{L_\mu},& \mu_{R}\rightarrow \omega^2 \mu_{R}, & \nu_{R_2}\rightarrow  \nu_{R_2},  \\ D_{L_\tau}\rightarrow \omega D_{L_\tau},& \tau_{R}\rightarrow  \omega \tau _{R}, & \nu_{R_3}\rightarrow \nu_{R_3}. \nonumber
\end{eqnarray}
With the complex scalar singlet $\chi$ transforming as $\chi \rightarrow \omega^2 \chi$ under $Z_3$, the type I seesaw contribution is given by
\begin{eqnarray} 
M_{\nu}^{I}= \left(
\begin{array}{ccc}
X& 0 &X \\
0 &0 &0 \\
X&0& X
\end{array}
\right)
\end{eqnarray}
and $Z_3$ invariant scalar triplet leads to the type II contribution given by
\begin{eqnarray} 
M_{\nu}^{II}= \left(
\begin{array}{ccc}
X& 0 &0 \\
0 &0 &X \\
0& X & 0
\end{array}
\right),
\end{eqnarray}
thus, yielding class $B_3$ of the FGM two zero texture neutrino mass matrices.
The required leptonic field transformations under $Z_3$ for class $B_4$ are given by
\begin{eqnarray}
 D_{L_e}\rightarrow D_{L_e},& e_{R}\rightarrow e_{R},&  \nu_{R_1}\rightarrow \omega \nu_{R_1},  \nonumber \\ D_{L_\mu}\rightarrow \omega D_{L_\mu},& \mu_{R}\rightarrow \omega \mu_{R}, & \nu_{R_2}\rightarrow  \nu_{R_2},  \\ D_{L_\tau}\rightarrow \omega^2 D_{L_\tau},& \tau_{R}\rightarrow  \omega^2 \tau _{R}, & \nu_{R_3}\rightarrow \nu_{R_3}. \nonumber
\end{eqnarray}
The complex scalar singlet $\chi$ is required to transform under $Z_3$ as $\chi \rightarrow \omega^2 \chi$ and the resulting type I contribution for this case is given by
\begin{eqnarray} 
M_{\nu}^{I}= \left(
\begin{array}{ccc}
X& X &0 \\
X &X &0 \\
0& 0 & 0
\end{array}
\right).
\end{eqnarray}
The $Z_3$ invariant scalar triplet leads to the type II contribution of the form
\begin{eqnarray} 
M_{\nu}^{II}= \left(
\begin{array}{ccc}
X& 0 &0 \\
0 &0 &X \\
0& X & 0
\end{array}
\right),
\end{eqnarray}
leading to class $B_4$ of the FGM neutrino mass matrices.\\
For classes $B_1$ and $B_2$, the leptonic fields are required to transform under $Z_3$ as
\begin{eqnarray}
 D_{L_e}\rightarrow D_{L_e},& e_{R}\rightarrow e_{R},&  \nu_{R_1}\rightarrow \nu_{R_1},  \nonumber \\ D_{L_\mu}\rightarrow \omega D_{L_\mu},& \mu_{R}\rightarrow \omega \mu_{R}, & \nu_{R_2}\rightarrow  \omega \nu_{R_2},  \\ D_{L_\tau}\rightarrow \omega^2 D_{L_\tau},& \tau_{R}\rightarrow  \omega^2 \tau _{R}, & \nu_{R_3}\rightarrow \omega^2 \nu_{R_3}, \nonumber
\end{eqnarray}
resulting in a diagonal Dirac neutrino mass matrix $M_D$. We do not require any scalar singlet to realize $M_R$. The resulting type I contribution becomes
 \begin{eqnarray} 
M_{\nu}^{I}= \left(
\begin{array}{ccc}
X& 0 &0 \\
0 &0&X \\
0& X & 0
\end{array}
\right).
\end{eqnarray}
However, in these cases the scalar triplet $\Delta$ transforms under $Z_3$ as
\begin{eqnarray}
\Delta \rightarrow\omega^2 \Delta(for class B_1)\\
\Delta \rightarrow\omega \Delta(for class B_2)
\end{eqnarray}
resulting in type II contribution of the form
\begin{eqnarray} 
M_{\nu}^{II}= \left(
\begin{array}{ccc}
0& X &0 \\
X &0 &0 \\
0& 0 & X
\end{array}
\right)
\end{eqnarray}
for class $B_1$ and
\begin{eqnarray} 
M_{\nu}^{II}= \left(
\begin{array}{ccc}
0& 0 &X \\
0 &X &0 \\
X& 0 & 0
\end{array}
\right)
\end{eqnarray} 
for class $B_2$. Type (I+II) seesaw yields classes $B_1$ and $B_2$ of the FGM two zero neutrino mass matrices. The scalar potential for these two cases is given by
\begin{eqnarray}
V(\phi, \Delta)=m^2\phi^\dagger\phi + M^2tr(\Delta^\dagger\Delta)+(\mu \phi^\dagger \Delta \tilde{\phi}+h.c.)+...,
\end{eqnarray}
where $\tilde{\phi}=i\tau_2\phi^*$ and the dots indicate quartic terms which respect $Z_3$. The dimension three term $\phi^\dagger \Delta \tilde{\phi}$ is not allowed in the scalar potential under $Z_3$. However, to obtain a small non-zero VEV of the scalar triplet \cite{14} we softly break the $Z_3$ symmetry by including this term in the scalar potential.\\
Class $C$ of FGM two zero textures cannot be realized by $Z_3$ symmetry, and the next higher order group $Z_4$ is used to realize this texture. The leptonic fields are required to transform under $Z_4$ as
\begin{eqnarray}
 D_{L_e}\rightarrow D_{L_e},& e_{R}\rightarrow e_{R},&  \nu_{R_1}\rightarrow \nu_{R_1},  \nonumber \\ D_{L_\mu}\rightarrow i D_{L_\mu},& \mu_{R}\rightarrow i \mu_{R}, & \nu_{R_2}\rightarrow i \nu_{R_2},  \\ D_{L_\tau}\rightarrow -i D_{L_\tau},& \tau_{R}\rightarrow  -i \tau _{R}, & \nu_{R_3}\rightarrow -i \nu_{R_3}, \nonumber
\end{eqnarray}
leading to a diagonal charged lepton mass matrix, a diagonal Dirac neutrino mass matrix $M_D$ and non diagonal right-handed Majorana mass matrix $M_R$. The type I contribution in this case is given by
 \begin{eqnarray} 
M_{\nu}^{I}= \left(
\begin{array}{ccc}
X& 0 &0 \\
0 &0&X \\
0& X & 0
\end{array}
\right).
\end{eqnarray}
However, for symmetry realization of class $C$ we are constrained to introduce two scalar triplets which transform as
\begin{eqnarray}
\Delta_1 \rightarrow i\Delta_1\\
\Delta_2 \rightarrow-i \Delta_2,
\end{eqnarray}
under $Z_4$ leading to the type II contribution of the form
\begin{eqnarray} 
M_{\nu}^{II}= \left(
\begin{array}{ccc}
0& X &X \\
X &0&0\\
X& 0& 0
\end{array}
\right)
\end{eqnarray}
resulting in class $C$ of FGM neutrino mass matrices. Again, we must break $Z_4$ softly in the scalar potential to obtain small non-zero VEV of the scalar triplets. A detailed phenomenological analysis of the FGM two zero textures has been presented in \cite{3}. Class C of FGM textures was obtained from a horizontal symmetry $Z_4$ with a scalar sector comprising of SM Higgs doublet and three scalar triplets \cite{14}. Class C has, earlier \cite{19}, been obtained in a model based on non-Abelian group $Q_8$ with a much more cumbersome scalar sector with two doublets and four triplets. Both these scenarios have been discussed in the context of type-II seesaw.
\section{Symmetry Realization of One Zero Textures}
\begin{table}[h]
\begin{center}
\begin{tabular}{|c|c|c|}
\hline  Zero Texture & $D_{L_e}(e_r)$, $D_{L_\mu}(\mu_r)$, $D_{L_\tau}(\tau_r)$& $\Delta_2$ \\
\hline $M_{\nu_{11}}=0$ &$\omega^2$ , $\omega$ , 1 & $\omega$ \\
\hline  $M_{\nu_{12}}=0$  & $\omega^2$ , 1, $\omega$ &$ \omega^2$ \\
\hline  $M_{\nu_{13}}=0 $ &  $\omega^2$ , $\omega$ , 1  & $\omega^2$ \\
\hline  $M_{\nu_{22}}=0$ & 1, $\omega^2$ ,$ \omega $& $\omega$ \\
\hline $ M_{\nu_{23}}=0$  & $\omega$ , 1, $\omega^2$  & $\omega^2$ \\
\hline $ M_{\nu_{33}}=0 $ & 1, $\omega$, $\omega^2$  & $\omega$ \\
\hline
\end{tabular}
\caption{Leptonic field transformations under $Z_3$ for all six possible one zero textures.}
\end{center}
\end{table}
One zero texture neutrino mass matrices can be realized from the cyclic $Z_3$ symmetry by extending the Higgs sector. These textures are realized by extending SM with two scalar triplets, three right-handed neutrino singlets and a complex scalar singlet. Under $Z_3$ the complex scalar singlet transforms as $\chi \rightarrow \omega^2 \chi$, whereas one of the scalar triplets $(\Delta_1)$ remains invariant. The right-handed neutrino fields transform as 
\begin{eqnarray}
 \nu_{R_1}\rightarrow \omega \nu_{R_1},  \nonumber \\  \nu_{R_2}\rightarrow \nu_{R_2},  \\ \nu_{R_3}\rightarrow \nu_{R_3}. \nonumber
\end{eqnarray}
 The transformations of other fields for all classes of one zero textures in $M_{\nu}$ are given in Table (2). 
The scalar potential for the one zero textures is given by
 \begin{eqnarray}
 V(\phi, \Delta, \chi)=m^2\phi^\dagger\phi + \sum_{j,k=1}^{2}(M^2)_{jk}tr(\Delta_j^\dagger\Delta_k)+ \lambda \chi^ {*}\chi+(\mu \phi^\dagger \Delta_1 \tilde{\phi}+h.c.)+...,
 \end{eqnarray}
 To obtain a small non-zero VEV of the scalar triplet $\Delta_2$ we softly break the $Z_3$ symmetry by including dimension two terms in the scalar potential where $j\neq k$. A detailed phenomenological analysis of one zero textures can be found in \cite{2}.
\section{Conclusions}
In the flavor basis, the zero textures of the neutrino mass matrix are realized using the small Abelian cyclic symmetry group $Z_3$/$Z_4$. All the zero textures are realized within the context of type (I+II) seesaw mechanism which is natural to SO(10) GUTs. Since, only one Higgs doublet is used in the realization of these zero textures, the flavor changing neutral current interactions are naturally suppressed. We introduced three right-handed neutrino singlets and at the most two scalar triplets which acquire small nonzero VEVs. However, in some cases we,  introduced an additional complex scalar singlet which transforms nontrivially under $Z_3$ symmetry. For one zero textures and some of the two zero textures we softly break the relevant symmetry to obtain a small non-zero VEV of scalar triplet. The zero textures in the neutrino mass matrix remain stable under RG evolution as the symmetry realization involves only the SM Higgs doublet.  

\textbf{\textit{\Large{Acknowledgements}}}

The research work of S. D. is supported by the University Grants
Commission, Government of India \textit{vide} Grant No. 34-32/2008
(SR). R. R. G. acknowledge the financial support provided by the Council for Scientific and Industrial Research (CSIR), Government of India.

\end{document}